\documentclass[letterpaper,prb,preprint,showpacs,12pt]{revtex4}
\usepackage{graphicx}
\pdfoutput=1
\usepackage{Tomspeak}
\usepackage{hyperref}
\usepackage{color}
\newcommand{\bbC} {{\bf C}} 
\newcommand{\bbD}{{\bf D}}
 
\newcommand{\uQ} {{\underline Q}}
\newcommand{\bbuC}{\underline \bbC} 
\newcommand{\calE}{{\cal E}} \def\bbI{{\bf I}}
\newcommand{\etal}{{\em et al}}
\long\def\omitt#1{}

\begin{document}
\title{Singular electrostatic energy of nanoparticle clusters}
\author{Thomas A.\ Witten}
\affiliation{Department of Physics and James Franck Institute,
University of Chicago, Chicago, Illinois 60637}
\author{Nathan Krapf}
\affiliation{Department of Physics and James Franck Institute,
University of Chicago, Chicago, Illinois 60637}
\date{\today}

\begin{abstract}
The binding of clusters of metal nanoparticles is partly electrostatic.  We address difficulties in calculating the electrostatic energy when high charging energies limit the total charge to a single quantum, entailing unequal potentials on the particles.  We show that the energy at small separation $h$ has a strong logarithmic dependence on $h$.  We give a general law for the strength of this logarithmic correction in terms of a) the energy at contact ignoring the charge quantization effects and b) an adjacency matrix specifying which spheres of the cluster are in contact and which is charged. We verify the theory by comparing the predicted energies for a tetrahedral cluster with an explicit numerical calculation.  
\end{abstract}
 \maketitle
\section{Introduction} \label{sec_introduction}

\begin{figure}[tbh]
\includegraphics[width=.9\hsize]{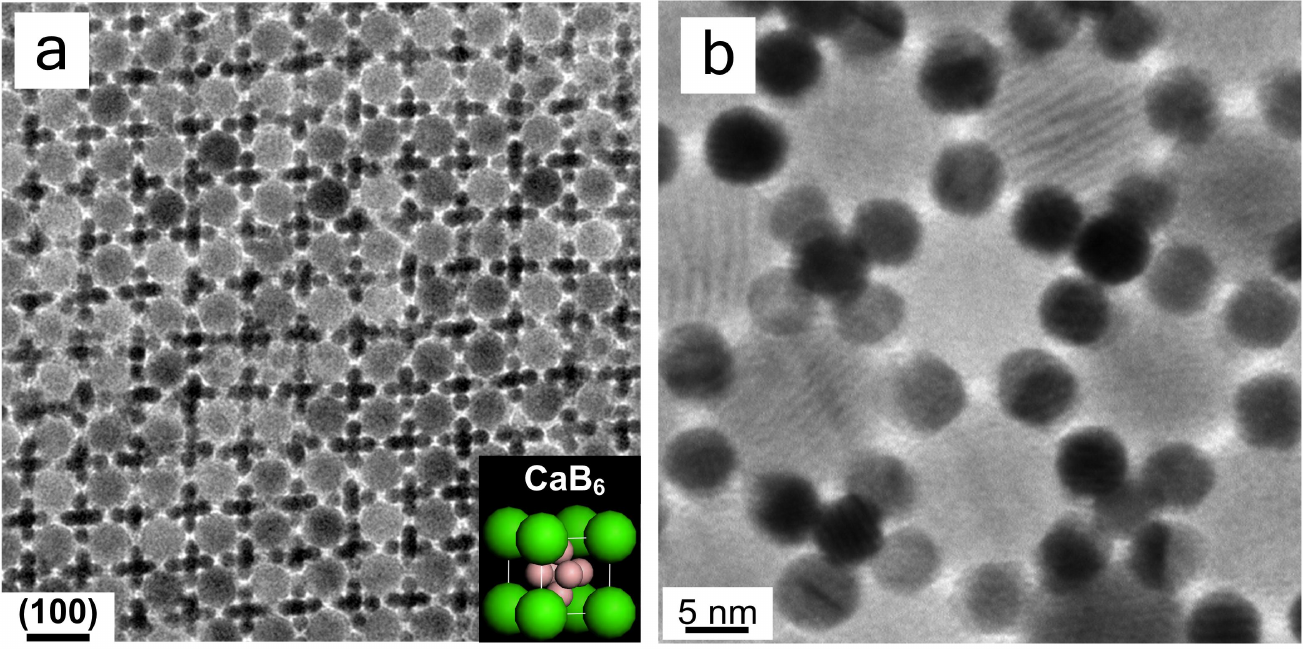}
\caption[]{ a: Transmission electron micrograph of experimental superlattice structure containing lead sulfate and dark colored palladium nanoparticles showing formation of regular palladium clusters as sketched in the colored inset.  Scale bar is 20 nm.  Reprinted by permission from Macmillan Publishers Ltd: from Ref. \cite{TalapinNature2006}, Figure 1j, courtesy D. V. Talapin. \hfill\break
b: Transmission electron micrograph of a dodecagonal quasicrystal superlattice self-assembled from Fe$_2$ O$_3$ nanocrystals and clustered dark-colored 5-nm gold nanocrystals.  Reprinted by permission from Macmillan Publishers Ltd from Ref. \cite{Talapin:2009uq} Figure 2b, courtesy D. V. Talapin.}
\label{fig_micrographs}
\end{figure}  

In self-assembled lattices of nanoparticles one often encounters clusters of metal particles\cite{Chen:2010fk}  as shown in Figure \ref{fig_micrographs}.  The remarkable stability of these clusters was argued to depend partly on states of nonzero electric charge\cite{TalapinNature2006}.  For particles of nanometer scale, such states are dominated by the quantization of charge.  The energy to add a single electron to a particle becomes large on the scale of the thermal energy $k_B T$, so that net charge on a particle is atypical.  Thus any net charge on a cluster is necessarily unevenly distributed over its particles. Still, a net charge on one particle must polarize the surrounding particles, producing electrostatic attraction.  This contrasts with the macroscopic case in which the available charge would be shared amongst the particles, producing repulsion.  It is of great interest to understand what types of clusters are favored under this simple and novel binding mechanism. Mutual electrostatic interactions between spherical conductors and with surfaces are of interest in space environments\cite{Boyer:1994kx} and in scanning probe microscopy\cite{Kalinin:2004rw}.  Merrill \etal \cite{Merrill:2009fk} explored the interactions among charged colloidal particles in clusters in solution. 

Unlike most interactions of small particles, this electrostatic interaction cannot be reduced to a pairwise potential energy. Charge on one sphere induces polarization on each nearby sphere. This polarization induces further polarization in other spheres. Since their separation is not large compared to their radius, the polarization cannot be accurately described by a dipole approximation.  Instead, all the spheres carry a polarization charge distribution that must be found self-consistently to minimize the electrostatic energy. It is not known what types of clusters would be favored by this novel multi-body interaction mechanism.   Recently Moore\cite{AMooreArXiV2010} has provided a multipole formalism for calculating this energy and has explored the energies of simple clusters .

These polarization effects are nontrivial even for the case of two isolated spheres.  Numerical solutions by A. Russell \cite{Russell:1923xe}, by Pisler \etal\cite{Pisler:1970zr} and by Kalinin \etal \cite{Kalinin:2004rw} have been developed. A case of special interest is that of identical spheres of radius $R$ bearing equal and opposite charge $q$ at separation $h$. At small separation $h \muchlessthan R$ the charge becomes concentrated arbitrarily strongly near the contact point. This concentrated contact charge creates a logarithmically singular mutual capacitance\cite{Jackson} $c(h)$ of the form $c(h) \goesto \frac 1 4 R \log(\alpha R/h)$, where $\alpha$ is a numerical constant.  The resulting electrostatic energy $\half q^2/c(h)$ reflects this singular behavior.  This divergent contact charge complicates the treatment of clusters of spheres with different charges. Moore's recent work on such clusters\cite{AMooreArXiV2010} shows a non-regular dependence of the energy on separation.  

\begin{figure}
\hbox to \hsize {\hfill 
\includegraphics[width=.4\hsize]{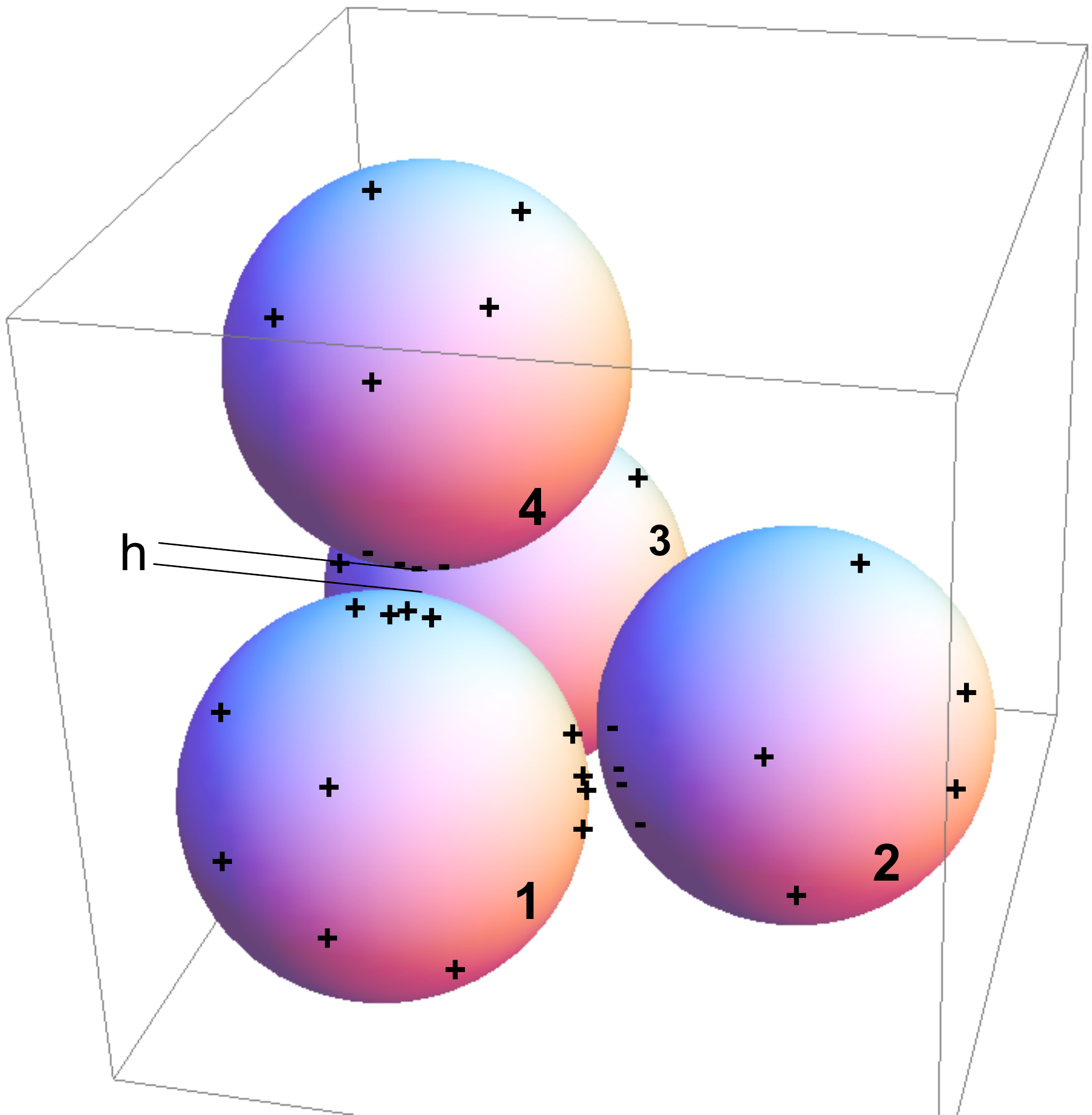}\hfill
\includegraphics[width=.4\hsize]{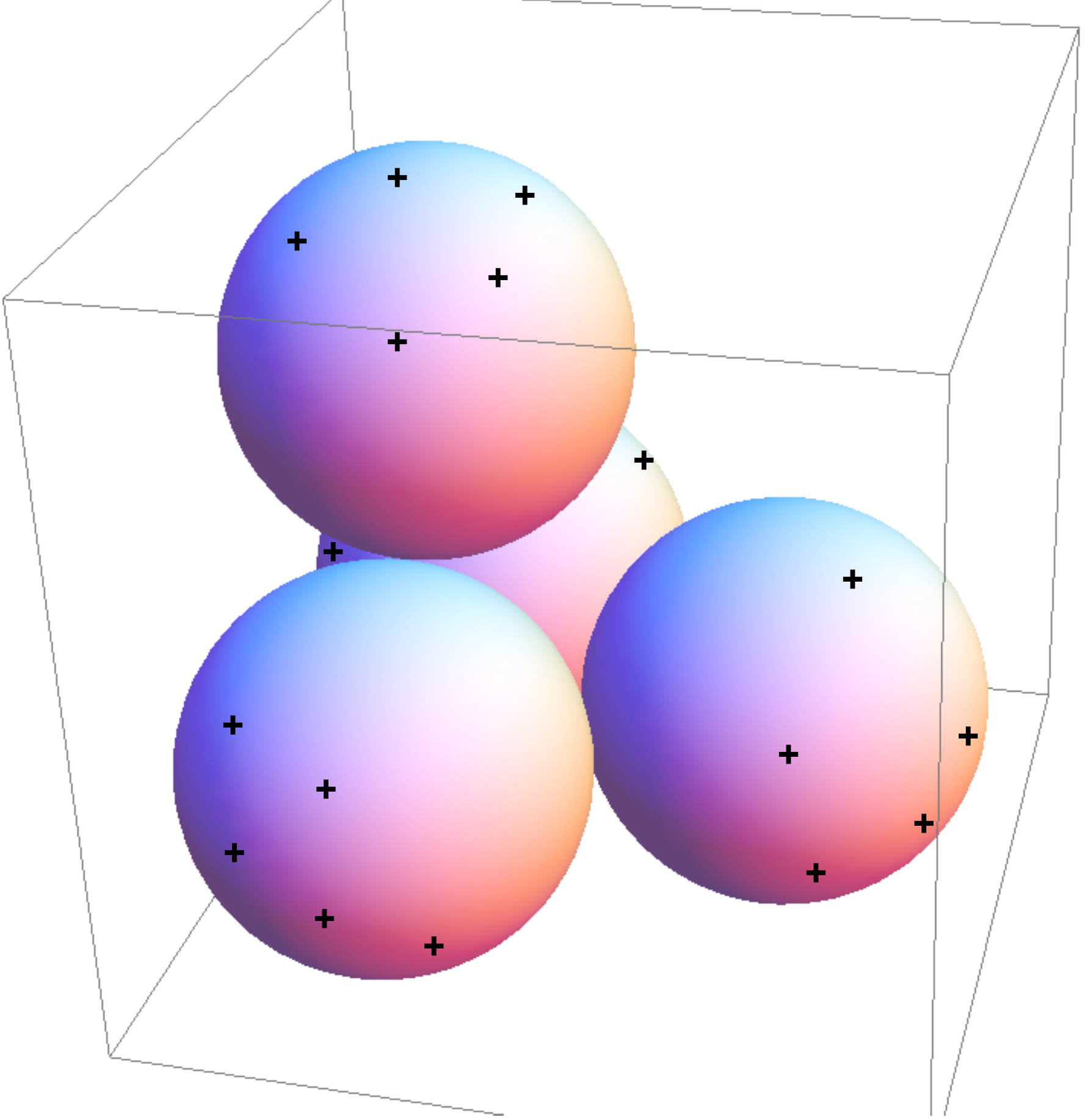}\hfill
}
\caption{Sketch of a cluster whose electrostatic energy is to be calculated. Sphere 1 has a net charge $Q$; spheres 2, 3, and 4 have no net charge.  Spheres 1, 2 and 3 are in near contact. Sphere 4 is in near contact with spheres 1 and 3 only.  Near contacts have separation $h$ much smaller than the sphere radii. Contact charges $q$ between spheres 1 and 2, and spheres 1 and 4 are shown.  Similar contact charges between spheres 1 and 3 and spheres 2 and 3 are hidden from view.  Right. Same cluster with the charge $Q$ free to migrate between spheres.  Spheres are at the same potential and there is no contact charge.  The regular tetrahedron treated in Figure \ref{fig_graph} is obtained by moving sphere 4 so that it contacts all the other spheres. }
\label{fig_sketches}
\end{figure}

Here we investigate the implications of the singular contact charge for the electrostatic energy $\calE(h)$ of clusters of conducting spheres $i$ at small separation $h$ when the total charge $Q$ resides on only one sphere, as sketched in Figure \ref{fig_sketches}.  
We contrast this energy to the simpler equipotential case where the charge $Q$ is allowed to pass freely between the spheres.  Here there is no contact charge, and the electrostatic energy $\calE_e(h)$ varies smoothly with $h$.   However in the case of interest where only one sphere is charged, new behavior arises owing to the appearance of contact charge. It is necessary to characterize this new behavior in order to find the desired electrostatic energy when the separations $h$ are small. We find that the energy at contact remains finite and equal to $\calE_e(0)$, but it acquires a logarithmic correction in $h$:
\begin{equation}
\calE(h)  \longrightarrow  \calE_e(0)~[1 + A/c(h) + ... ] .
\label{eq_announce_result}
\end{equation} 
Remarkably, $A$ depends only on the equipotential charge distribution and an adjacency matrix of the cluster considered.  

We begin by reviewing the origin of the singular $c(h)$.  Next we define a capacitance matrix $\bbC(h)$ that gives the proportionality between the charges $Q_i$ and the potentials $V_i$.  We then isolate the singular contribution to $\bbC(h)$ and thereby derive the result of Eq. \ref{eq_announce_result}.    To gauge the importance of the logarithmic correction in practice, we work out the case of a tetrahedral cluster.  Finally we comment on experimental implications and tests.
 
\section{Mutual capacitance of two spheres near contact}
\label{sec_twospheres}

For completeness we recall the origin of the logarithmic divergence of the mutual capacitance of two neighboring spheres of radius $R$, bearing equal and opposite charges $Q$.  The potential difference between the spheres is denoted $V$.  In the limit $h/R \muchlessthan 1$, the capacitance is dominated by the adjacent sections of the two spheres.  Since the curvature there is very small on the scale of $h$, we may find the capacitance from this region via the Derjaguin approximation\cite{Safran:1994uq}.  This approximation treats the system as a set of concentric annular ring capacitors, neglecting the slopes of the surfaces within each ring.  At lateral distance $x$ from the central axis, the electric field $E$ is evidently $V/y(x)$, where $y(x)$ is the gap thickness at position $x$.  Thus the surface charge density $\sigma(x) = E/(4\pi) = V/(4 \pi y(x))$.  To find the charge $Q$, we integrate $\sigma$:
\begin{equation}
Q = \integral 2\pi~x~dx \sigma(x)
\end{equation}
We note that the local height $y(x)$ is given by $x^2 + (R - (y - h)/2)^2 = R^2$ so that $2x~dx +  2 (R - (y - h)/2)~(-\half )dy = 0$ or for $R \muchgreaterthan y$,
\begin{equation}
2x~dx \goesto R~dy, 
\end{equation}
and
\begin{equation}
Q \goesto \integral_h^{a R}  2\pi V~ R~ dy / (4\pi y), 
\end{equation}
where $a~ R$ is some upper cutoff of thickness where the Derjaguin approximation breaks down.
 Thus,
\begin{equation}
Q \goesto \half V~R \integral_h^{a~R} dy/y = \half V~R~ \log(a~ R/h)
\end{equation}
as claimed.  The capacitance $c(h)=Q/(2V)$ thus goes logarithmically to infinity as $h\goesto 0$. Numerically, one can use image charge methods\cite{Kalinin:2004rw} to determine the explicit form of $c(h)$ for small $h$: $c(h) \aboutequal \frac 1 4 R\log (1.26 R/h)$.

From this capacitance we can infer the energy needed to separate the contacting spheres with charges $\plusorminus Q$.  At contact, the energy $\calE(0)$ is given by $\half Q^2/C$.  Since $C \goesto \infinity$, we have vanishingly small $\calE$ at contact.  At infinite separation we have the full Coulomb self energy $2 ~\half Q^2/R$.  Thus with equal and opposite charges the polarization of the spheres cancels virtually all the electrostatic energy of the separated spheres.

\section{Equipotential cluster}
\label{sec_equipotential}

We now consider a cluster of $n$ spheres labeled by $i$. We denote the set of charges $Q_i$ by the vector $\vector Q$. There is in general a linear relationship between the charges $\vector Q$ and the potentials on the spheres $\vector V$ of the form 
\begin{equation}
\vector Q = \bbC \vector V, 
\end{equation} 
where $\bbC$ is an $n\times n$ symmetric matrix.
 To set the stage for the problem of interest, we first consider the simpler problem of an equipotential cluster at potential $V_e$.  It is convenient to define a ``uniform vector'' $\vector u \definedas (1, 1, 1, .... 1)$.  Then in this equipotential case, the potentials have the form $\vector V \definedas  V_e \vector u$. Likewise, the total charge $Q$ is given by $\vector Q\cdot \vector u$.  The equipotential cluster has none of the contact charge considered above, and thus none of the non-regular dependence on separation $h$.  That is, $\bbC(h)  \vector u$ may be replaced by $\bbC(0) \cdot \vector u$. Thus the charges $Q_i$ also depend smoothly on $h$. For a given total charge $Q$, the potential $V_e$ is then given using $Q = \vector u \cdot \vector \uQ = V_e ~\vector u \cdot \bbC \vector u$. Evidently the equipotential capacitance $C_e$ is simply $Q/V_e = \vector u \cdot \bbC \vector u$.  
 
\section{Asymmetrically charged cluster}
\label{sec_assymetrically}

We now fix the charge $Q_i$ on each sphere. For definiteness we may consider the case where sphere 1 has charge $Q$ and the others are uncharged. We seek the corresponding potentials $\vector V$ that create this charge distribution.  From this case, one may infer the case of general $Q_i$ by superposition.   In general the potentials $V_i$ are now unequal.  The linear relationship $\vector Q = \bbC \vector V$ now includes contact charge and hence a non-regular dependence on separation $h$.   

We note that two isolated spheres $i$ and $j$ at separation $h$ much less than the radius have charges $q_{ij}$ given by $c(h) (V_i - V_j)$.  For our cluster of spheres, we define the contact charge between two spheres as this $q_{ij}$.  Thus the total contact charge $q_i$ on a sphere can be written $\sum_j q_{ij} = c(h) \bbD \vector V$.  Here the adjacency matrix $\bbD_{ij}$ is -1 for contacting spheres $i$ and $j$, while $\bbD_{ii}$ is the number of spheres contacting sphere $i$.  We note that for $n$ connected spheres this $\bbD$ has at least one null vector, since any uniform potential (proportional to the uniform vector $\vector u$) produces no contact charge.  Likewise the range space of possible contact charges $\vector q$ has at least one constraint: the sum of all the $q_i$ \ie $\vector u \cdot \vector q$ is just the sum of the $q_{ij}$, which vanish pairwise.  Thus $\vector u \cdot \vector q = 0$.  

Within these restrictions the contact matrix $c \bbD$ is invertible\cite{Bohar1991}.  That is, every $\vector q$ with $\vector u \cdot \vector q = 0$ has a corresponding potential $\vector v$ such that $\vector q = c \bbD \vector v$ and $\vector v \cdot \vector u = 0$. The Appendix establishes this invertibility property.  

Not all the charge in the sphere cluster is contact charge.  Thus in general for a given set of potentials $\vector V$ the total charge $\vector Q = \bbC \vector V$ has the form $\vector q + \vector \uQ$.  That is
\begin{equation}
\bbC \vector V = c(h) \bbD \vector V + \vector \uQ
\end{equation}
Evidently $\vector \uQ$ can be written in the form $\bbuC \vector V$; here $\bbuC$ is the capacitance matrix for the non-contact charge.  We suppose that only the contact charge $\vector q$ diverges for given $\vector V$ as $h \goesto 0$.  That means $\bbuC(h)$ is regular as $h \goesto 0$.  Thus for $h \muchlessthan R$ we may replace $\bbuC(h)$ by $\bbuC(0)$.

We now note that the singular behavior of contact charge implies that $\vector V$ becomes uniform as $c(h) \goesto \infinity$, independent of $\vector Q$.  
\begin{equation}
\vector Q = \bbuC \vector V + c(h) \bbD \vector V
\end{equation}  
When we let $c \goesto \infinity$ for fixed $\vector Q$,  any nonuniform $\vector V$ has a nonvanishing $\bbD \vector V$, and thus $c(h) \bbD \vector V$ diverges.  Such a divergence is incompatible with the fixed limiting values of $\vector Q$ and $\bbuC \vector V$.  Thus $\vector V$ cannot have a nonuniform part; that is $\vector V = V_e \vector u$ for some $V_e$. Recalling the discussion of equipotential clusters above, $V_e$ is the potential of the cluster with the same total charge as the desired cluster, with the spheres all at the same potential.

\section{Correction for noninfinite $c(h)$}
\label{sec_correction}

For large $c(h)$ the departure of $\vector V$ from the uniform $V_e \vector u$ must be small.  Accordingly we express $\vector V  \definedas V_e \vector u + \vector v$, where $\vector v$ is a small correction.  Using this form
\begin{equation}
\vector Q = \bbuC (V_e \vector u + \vector v) + c(h) \bbD \vector v
\label{eq_bbuC}
\end{equation}  
In this expression $\bbuC \vector u$ may be simplified.  This quantity is simply the non-contact charge under a uniform potential $V_e$.  However in this equipotential situation {\it all} the charge is noncontact charge, so that $\bbuC \vector u$ is simply $\bbC \vector u$.  Using this simplification, we obtain an implicit expression for $\vector v$:
\begin{equation}
\vector X \definedas \vector Q - V_e~ \bbC \vector u = \Big(\bbuC + c(h) \bbD \Big)
\vector v 
\label{eq_vectorv}
\end{equation}
We note that the charge vector $\vector X$ on the left has a vanishing total charge $\vector u \cdot \vector X$:
\begin{equation}
\vector u \cdot \vector X = \vector u \cdot (\vector Q - V_e~ \bbC \vector u)
= Q - V_e \vector u \cdot \bbC \vector u
\end{equation}
Since $V_e$ is defined by $Q = V_e \vector u \cdot \bbC \vector u$, the total charge in $\vector X$ vanishes as claimed.

Eq. \ref{eq_vectorv} expresses the desired $\vector v$ in terms of the regular matrices $\bbuC$ and $\bbD$, which have no singular dependence on $h$. In order to find $\vector v$ from this implicit expression, we must invert the matrix $\bbuC + c(h) \bbD$.  
We discuss the invertibility of $\bbD$ in the Appendix.  There we show that $\bbD^{-1} \vector X$ is well-defined for any $\vector X$ with $\vector u \cdot \vector X = 0$.   Moreover, the desired matrix $\bbuC + c(h) \bbD$ is invertible apart from exceptional $c(h)$ values, as the Appendix also explains.

We now consider the leading correction for small $h$, \ie\  large $c(h)$. In view of the last term in Eq. \ref{eq_bbuC}, $\vector v$ must be of order $1/c(h)$.  To order $(1/c(h))^0$, the equation reads.
\begin{equation}
\vector Q - V_e~ \bbuC \vector u = c(h) \bbD \vector v
\end{equation}  
Recalling that $\bbuC \vector u = \bbC \vector u$, we now obtain the potential correction $\vector v$:
\begin{equation}
\vector v = (1/c(h)) \bbD^{-1} (\vector Q - V_e~ \bbC \vector u)
\end{equation}
Since $\vector v$ is now determined, the state of the system for small separation $h$ is determined.  

As anticipated, the potential is nearly that of a cluster with only the total charge $Q$ constrained.  To determine the logarithmic correction owing to non-infinite $c(h)$ it suffices to know a) the adjacency matrix $\bbD$ and b) the charge distribution for the equipotential cluster, namely $\bbC \vector u$.  

As $h$ increases, further corrections become important.  Further calculation via this scheme would require $\bbC \vector V$ for nonuniform $\vector V$'s.  Such extensions would be cumbersome to implement.  However, even without carrying out such an expansion, the adjacency matrix $\bbD$ can be used to produce a smoother behavior in $h$.  If one has a scheme for computing $\bbC(h)$ for some range of $h$, then one may remove the singular behavior induced by $c(h)$ by constructing $\bbuC(h) \definedas \bbC(h) - c(h)\bbD$.  This smooth $\bbuC(h)$ can be used to find quantities of interest for values of $h$ where $\bbC(h)$ itself would be poorly converged.

From this $\vector v$ the electrostatic energy may readily be found, as we now show.  

\section{Electrostatic energy}
\label{sec_energy}

Given this expression for the potential vector $\vector V$, we may find the electrostatic energy $\calE$ for the cluster with a single charged sphere.  This $\calE$ can be expressed as $\half \sum Q_i V_i = \half \vector Q \cdot \vector V$.  For the case of interest, only sphere 1 is charged.  For convenience we define the vector $\vector 1 \definedas  (1, 0, 0, 0, ... 0)$.  Now the energy $\calE = \half  Q~ \vector V\cdot \vector 1$, that is,
\begin{equation}
{\calE} = \half Q (V_e + \vector v \cdot \vector 1) .
\end{equation}
To leading order this energy is simply $\half Q^2/(\vector u \cdot \bbC \vector u )$ or $\half Q^2/C_e(0)$, Here $C_e(0)$ is clearly larger than the capacitance  $C_1$ of an individual sphere.  Thus $\calE$ is smaller than that of the separated spheres, \viz\ $\half Q^2/C_1$.  Indeed, for a compact cluster of spheres, the capacitance is proportional to the radius, and hence to the 1/3 power of the number of spheres.  For a cluster of many spheres, the binding energy of the cluster approaches the electrostatic energy of the charged sphere, as in the two-sphere case with zero net charge treated above.

To evaluate the first correction in $\vector v$, we may express $\vector v$ in terms of the uniform potential $V_e$ using the asymptotic formula of Eq. \ref{eq_vectorv}.
\begin{equation}
\vector v\cdot \vector 1 \goesto (1/c(h))~ 
\vector 1\cdot \bbD^{-1}(Q ~\vector 1 - V_e~ \bbC \vector u)  
\end{equation}
Recalling that $V_e = Q/C_e(0)$, this yields
\begin{equation}
{\calE} \goesto \half~ {Q^2 \over C_e(0)} \left (1 + 
{C_e(0) \over c(h)} ~ 
\vector 1\cdot \bbD^{-1}\left [\vector 1 - {\bbC \vector u \over \vector u \cdot \bbC \vector u} \right ]   \right ) .
\label{eq_calEanswer}
\end{equation}

	This energy has the form announced in Eq. \ref{eq_announce_result}; evidently the part $(1 + A/c(h))$ in Eq.  \ref{eq_announce_result} is the quantity in $(... )$ above. 
	
This expression shows that one may isolate the singular part of the electrostatic energy for a cluster of conducting spheres close to contact, using nonsingular quantities which can be readily computed numerically.  The energy at $h=0$ but without conductance between spheres is the same as for the equipotential case where conductance is allowed.  This means that the imposed distribution of charge among the spheres has no effect on the energy at $h=0$.  

Increasing $h$ necessarily increases the energy.  Thus the correction in $1/c(h)$ is necessarily positive  (though this is not obvious from Eq. \ref{eq_calEanswer}).  As seen from Figure \ref{fig_graph} below, this increase can depend strongly on $h$. In general it depends on which sphere is charged.  One expects that the increase with $h$ is strongest for spheres that have many adjacent neighbors.  

\section{Numerical illustration}
\label{sec_illustration}
\begin{figure}
\includegraphics[width=\hsize]{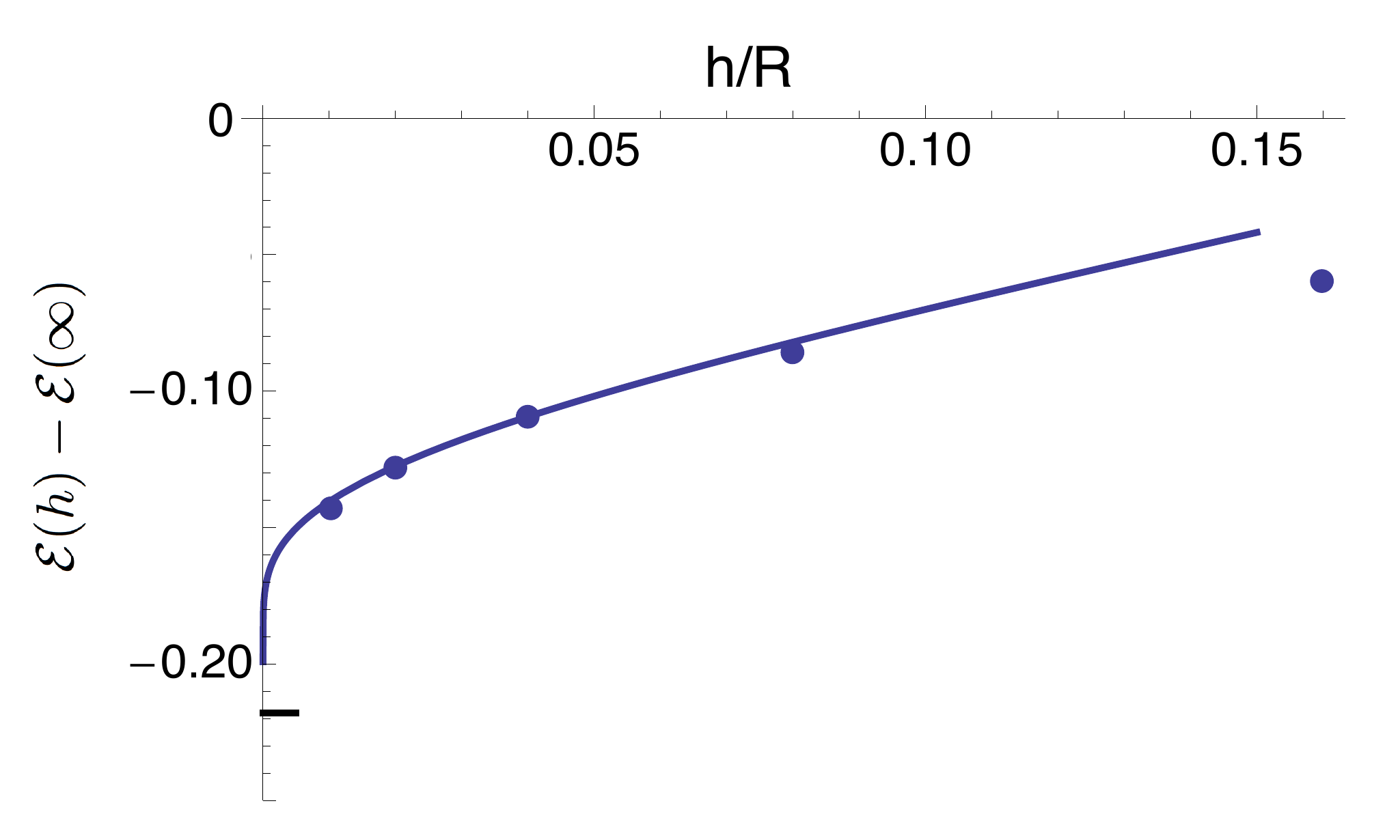}
\caption{Binding energy of a tetrahedron of unit spheres where one has unit charge. Units\cite{Jackson} are such that an isolated sphere has $\calE=\half$.  Points are from numerical energy minimization with charge represented as point charges, as
 described in the text.  Left two points used 1999 points per sphere; remaining points used 499 points per sphere.  Curve was obtained from Eq \ref{eq_calEanswer}.  The $\calE_e(0)$ needed in this equation was obtained from the method of Moore\cite{AMooreArXiV2010} and verified by the point charge method. The curve has no adjustable parameters.  This $\calE_e(0)$ is shown by the heavy mark on the vertical axis. For 5nm diameter spheres\cite{TalapinNature2006} in vacuum with a single electron charge, this binding energy is 5.0 kT at room temperature.}
\label{fig_graph}
\end{figure}
\omitt{ calculation of real-world binding energy see Figure1.nb
radius: 5nm/2 per Talapin nature article
charge = e = 4.8 E-10 statCoulombs = 1.6E-19 coul. (3 E9 statcoul/coul)}

In order to demonstrate the behavior of the formula in practice, we show a specific example:  a regular tetrahedron of contacting spheres, with a charge $Q$ on one of them.  Since each sphere contacts all three others, the adjacency matrix $\bbD$ is given by  
\begin{equation}
\bbD = \left [\matrix
{3 & -1 & -1 & -1 \cr
-1 & 3 & -1 & -1 \cr
-1 & -1 & 3 & -1 \cr
-1 & -1 & -1 & 3 \cr} \right ]
\end{equation}

The equipotential charges $\bbC \vector u$ are evidently all equal.  The capacitance $C_e(0)$ of the equipotential cluster is greater than that of a single sphere.  We report its numerical value below.  Evidently the ratio $\bbC \vector u / (\vector u \cdot \bbC \vector u) =  (1/4, 1/4, 1/4, 1/4)$, and $\vector 1 - \bbC \vector u / (\vector u \cdot \bbC \vector u) = (3/4, -1/4, -1/4, -1/4)$.  In order to compute $\vector 1 \cdot\bbD^{-1}(~\vector 1 - \bbC \vector u / (\vector u \cdot \bbC \vector u)~)$, we must find the vector $\vector w$ which solves $\bbD \vector w = (3/4, -1/4, -1/4, -1/4)$ with $\vector w \cdot \vector u = 0$.  The solution is 
\begin{equation}
\vector w = (3/16, -1/16, -1/16, -1/16).
\end{equation}
Finally, 
\begin{equation}
\vector 1 \cdot \bbD^{-1}(\vector 1 - \bbC \vector u / (\vector u \cdot \bbC \vector u)~) =  \vector 1\cdot \vector w = 3/16.
\end{equation}

 \par

To test the range of validity of Eq. \ref{eq_calEanswer}, we numerically evaluated the energy $\calE$ for various separations $h$.  We placed $N$ point charges of variable strength over the surface of each sphere and numerically adjusted these charges to minimize the mutual Coulomb energy of the charges under the constraint that the total charge on sphere 1 should be 1 while the others are zero.  Separately, we calculated this energy with only the total charge on all spheres constrained to be 1.  This gave an electrostatic energy at contact of 0.281.  The binding energy of the charged sphere to the others is thus evidently $0.281 - \half$.  We confirmed this value separately using the multipole method of Moore\cite{AMooreArXiV2010}


The dependence of $\calE$ on $h$ is shown in Figure 1.  The prediction of Eq. \ref{eq_calEanswer} is accurate at the ten percent level out to separations $h$ of roughly $0.1 R$ in this case.  

\section{Discussion}
\label{sec_discussion}

The preceding sections have explored a peculiar type of Coulomb interaction arising from the charging constraints encountered at the spatial scales of nanoparticles.  Below we note the limitations of our work and suggest experimental situations where the interaction discussed here might nevertheless be relevant.  

In order to demonstrate the specific features our mechanism, we have considered the simplest example that shows the necessary features.  First, a cluster of spheres like those considered here has charge polarization extending beyond the induced dipoles normally considered.  Second, any excess charge on the cluster is dominated by single electron charges residing on one or another of the particles.  Given these two features, one should observe the singular dependence on separation $h$ found above. Our main aim has been been to show the form of this singularity and how its magnitude may be predicted from simple geometric information.  The relative electrostatic energies of different clusters are important for determining their relative abundance and stability.  For real experimental situations  the relative abundance of actual cluster shapes doubtless depends strongly on several other factors as well.

In real clusters, it is artificial to assume a single charge on a particular particle; a number of charge distributions likely have significant probability.  If a single charge is present, it may reside on any particle of a cluster that doesn't require an extra energy much higher than $k_B T$.  Thus in practice one may need to consider an average over several charge positions in order to determine the stability of a given cluster shape.

In real materials a net charge on a cluster is only created in combination with a countercharge elsewhere.  In situations where the charging energy is large on the scale of $k_B T$, these countercharges are normally located close to their opposites.  In addition one expects ambient charges of both signs; these ambient charges screen the potential arising from any assumed net charge.  Either the countercharge or the screening effects would greatly alter the relative energies of different clusters.   
Yet these effects of other charges need not be crucial.   As for the countercharge effect, one may sequester the countercharge so that it remains far from the cluster in question.  One example of such sequestration is in the layered semiconductors used to make two-dimensional electron gases\cite{Pfeiffer:1990ys}.  Here the countercharge consists of strong electron donors or acceptors held many nanometers away from the free charges of interest.  The same effect is achieved by making the counterions reside in large colloidal particles or micelles\cite{Sainis:2008uq}.  Their large size entails a low surface potential and hence weak interaction with the charges of interest.  In these same colloidal systems, the high charging energy assures that ambient charge is minimal so that screening is not important.

Naturally real clusters like those of Figure \ref{fig_micrographs} experience other forms of interaction unrelated to net charge on the cluster.  The organic coronas\cite{TalapinNature2006} used to to stabilize the particles exert interparticle forces.  So do steric interactions with other neighboring nanoparticles.  Dispersion forces and solvent-specific chemical interactions are also present.  In order to make reliable predictions of cluster shapes, one would need to add these conventional interactions to the charge-induced interactions considered here.  

Experimental consequences of our clustering mechanism could potentially be found in the binary lattices like Figure 1 that motivated our study.  If our mechanism is important, one expects a) cluster shapes with lower electrostatic energy as calculated above should be relatively more prevalent, and b) particles with a thicker ligand layer should be less strongly bound but have greater preference for specific charge sites.  Still, the number of competing effects that determine the specific cluster shapes precludes any decisive predictions. 

Other simpler systems give a brighter prospect for decisive predictions.  
One such system is a dilute dispersion of nanoparticles in a nonpolar solvent\cite{Fernandez:2009fk}.  One may induce charge separation by adding large counterions to the dispersion\cite{Sainis:2008uq}.  Then any nanoparticles with a net charge will attract neutral nanoparticles via the mechanism described above.  If the counterions are sufficiently large and distant, their effects can be made minor.  Then one expects to observe clusters with relative abundance dictated by the electrostatic binding energies described above.  

\section{Conclusion}
\label{sec_conclusion}  

As shown above, the electrostatic energy of a cluster of spherical conductors has a novel form when one conductor is charged and their separations are small.  In the limit of small separations the energy is finite, but the corrections to this limit are logarithmically singular.  Thus for real clusters where the separation is nonzero, it is important to know the singular contribution.  Both the limiting energy and the corrections can be expressed in terms of non-singular operations. It appears from our numerical example that these small separations can have a significant impact on the binding of the clusters. In certain situations as noted above, this distinctive form of binding could be significant in determining the prevalent cluster shapes. 

With these methods in hand, one may readily compare the electrostatic binding energies of a range of cluster shapes.  Our work on these comparisons is in progress. 

\begin{acknowledgments}
The authors are grateful to Prof. Dmitri Talapin for numerous discussions of his experimental findings on nanoparticle self-assembly.  We thank Toan Nguyen, Eric Dufresne and Jason Merrill for helpful discussions.   We thank Alexander Moore, author of Ref. \cite{AMooreArXiV2010}, for extensive discussions and the use of his algebraic code. We thank the Aspen Center for Physics for hospitality during the completion of this work.  
This work was supported by the National Science Foundation's MRSEC
Program under Award Number DMR 0820054   
\end{acknowledgments}

\appendix* 
\section{Existence of inverses in Eq. \ref{eq_vectorv}}

Our general equation for the nonuniform part of the potential $\vector v$ requires that the matrix inversion implicit in Eq. \ref{eq_vectorv} is well-defined.  As we have seen, the matrix $\bbD$ is not invertible as it stands.  Thus the invertibility needed for Eq. \ref{eq_vectorv} needs to be verified.  We first discuss the invertibility of $\bbD$ for the restricted $\vector v$'s and $\vector Q$'s whose uniform part vanishes ($\vector u \cdot \vector v = 0$) and whose total charge vanishes ($\vector u \cdot \vector Q = 0$). The $\bbD$ matrix is known in graph theory as the Laplacian Matrix\cite{Bohar1991}, and its invertibility properties are well established.  Here we explain these properties in the present context.   We then argue that $\bbuC + c(h) \bbD$ is also invertible for the purposes of Eq. \ref{eq_vectorv}.

To analyze $\bbD$, we picture a circuit of consisting of $n$ nodes connected by capacitors of capacitance $c$. Every contact point between two spheres $i$ and $j$ is represented by a capacitor connected between nodes $i$ and $j$.  Then the equations $\vector Q = c \bbD \vector V$ are simply the circuit conditions of equilibrium.   The capacitance matrix of this network is $c\bbD$.  The desired invertibility property is that for any $\vector Q$ with $\vector u \cdot \vector Q = 0$, there exists a $\vector v$ with $\vector u \cdot \vector v = 0$ such that $\vector Q = c\bbD~ \vector v$. 
Circuit theory\cite{CircuitTheory} gives conditions such that the charges $\vector Q$ specify the potentials $\vector v$.  Below we verify this for  any connected set of nodes.

For simplicity we consider a basis set of $\vector Q$ where an arbitrary node $k$ has charge 1 while node 1 has charge -1.  We first establish the result for a chosen subset of the final set of contacts and then show that it remains true when new contacts are added.  We choose the initial subset  to be a ``spanning tree"\cite{Bohar1991} such that there is only one contact path between any two spheres.  Then all spheres along the path between $k$ and 1 have contact charges $\plusorminus 1$ and all the potential differences from sphere to sphere along the path $k \goesto 1$ are equal to $-1/c$.
The remaining contacts, not on the $k\goesto 1$ connecting path, have no potential difference.  Thus any sphere $i$ not on this connecting path has the same potential as its connected neighbors.  Indeed, all the spheres $i$ that connect to a sphere $j$ on the $k\goesto 1$ connecting path share its potential $V_j$.  Thus for this case, there is a potential vector for any given charge vector.  The above argument only defines potential differences, Thus it specifies $\vector V$ only up to an additive constant. To define a unique $\vector V$, we specify further that the sum of the potentials must vanish.  Following our notation in the main text, we denote this restricted set of potentials as $\vector v$. This shows the desired result for the spanning tree network.

From this base we may inductively construct the new potentials that obtain after a single contact is added, \eg between spheres $i$ and $j$. Given a set of charges $\vector Q$ (summing to zero), we presume that these potentials $\vector x$ are determined in the given network without the $i$-$j$ contact. We then show that the $\vector v$ are also determined after the new contact has been added.  We denote the difference $x_i - x_j \definedas A $. It is useful to define the potentials for a particular $\vector Q$, namely, a charge $Q'$ at $i$ and $-Q'$ at $j$. This charge necessarily resides on the existing contacts of sphere $i$ and sphere $j$.  The corresponding set of potentials is proportional to $Q'$; we denote them as $\vector w~ Q'$.  We denote the potential difference between $i$ and $j$, \ie $(w_i - w_j) Q'$, as $B ~Q'$.  

We now add the new contact, and claim that the desired potential $\vector v$ is a linear combination of $\vector x$ and $\vector w$.  This potential must satisfy the network equation for the new contact as well as those of the pre-existing contacts.  Both $\vector x$ and $\vector w Q'$ satisfy these equations for the pre-existing contacts.  When the new contact is added, a new contact charge $\tilde q$ appears on sphere $i$ and $-\tilde q$ appears on $j$. The desired $\vector v$ must satisfy $\tilde q = c (v_i - v_j)$.  However, the new contact must not change the total charge on $i$ or $j$. Thus the pre-existing contacts on node $i$ must change their total charge by an amount $Q' = - \tilde q$. The new network with the added contact must have the same potentials as the pre-existing network with the added charges $Q'$.  
This $Q'$ generates the potentials $\vector w Q' = -\vector w \tilde q $.  Thus the net potential vector is $\vector x - \vector w \tilde q$.  The potential difference between $i$ and $j$ is then $A - B \tilde q$.  The contact equation for the new contact then requires $c \tilde q = A - B \tilde q$ \ie\ $\tilde q ~= c A / (1 + cB)~$.  Moreover, the corresponding potential vector $\vector v = \vector x -\vector w ~\tilde q$ satisfies both the new and pre-existing contact equations with the original charges $\vector Q$. This shows that the charges determine the potentials in the new network, as claimed.  Since $\vector q$ in this system depends only on potential differences,  we may readily add a constant to all the potentials to assure that $\vector u \cdot \vector v = 0$.

Our general expression for determining $\vector v$ depends on inverting the matrix $\bbuC + c(h)\bbD$.  This expression is our decomposition of the full capacitance matrix $\bbC$, which we presume to be invertible. In order to be useful, this inverse must be readily calculable for generic values of $c(h)$.  We now show that this is true provided $\bbuC$ is invertible. We may re-arrange the equation $(\bbuC + c(h)\bbD) ~\vector v = \vector Q$ to form
\begin{equation}
\left (\bbI + (1/c) \bbuC \bbD^{-1}\right )\bbD \vector v = \vector Q
\end{equation}
where $\bbI$ is the identity matrix.  The matrix in $(....)$ is invertible (for $\vector u \cdot \vector Q = 0$) for generic values of $c$.  To show this, we first note that $\bbuC \bbD^{-1}$ is invertible, since its two factors are.  Further, since the two factors are symmetric matrices, the product $\bbuC \bbD^{-1}$ is also symmetric.  Thus $\bbuC \bbD^{-1}$ has a simple eigenvector expansion with no zero eigenvalues. We denote these eigenvalues as $\lambda_i$.  The full expression $(...)$ also has a full eigenvector expansion with eigenvalues $1 + \lambda_i/c$.  Provided none of the $\lambda_i$ are equal to $-c$, all these eigenvalues are also nonzero. Since   $\bbI + (1/c) \bbuC \bbD^{-1}$ has a complete set of nonzero eigenvalues, it too is invertible, as claimed. Then we can determine $\vector v$ by
\begin{equation}
\vector v = \bbD^{-1} \left (\bbI + (1/c) \bbuC \bbD^{-1}\right )^{-1} \vector Q
\end{equation}

\end{document}